\documentclass{article}
\usepackage{fancyhdr}
\usepackage{paralist}
	\usepackage{graphics}
	\usepackage{epsfig}
	\usepackage{graphicx}
	\usepackage{epstopdf}
\usepackage[colorlinks=true]{hyperref}
\usepackage{epsfig}
\usepackage{amsmath,amssymb,tabularx}
\usepackage{enumitem}
\usepackage{caption,subcaption}
\usepackage[usenames,dvipsnames,svgnames,table]{xcolor}
\usepackage[a4paper,top=2cm,bottom=2cm,left=2cm,right=2cm]{geometry}
\hypersetup{urlcolor=blue, citecolor=red}

\usepackage{ulem}

\newcommand{\f}{\mathbf{f}}

\graphicspath{{./}{figure/}}

\newcommand{\cT}{\mathcal{T}}
\newcommand{\cF}{\mathcal{F}}

\newcommand{\vf}{\varphi}

\newcommand{\ve}{\varepsilon}

\newcommand{\bnu}{\boldsymbol{\nu}}

\newcommand{\bx}{{\boldsymbol{x}}}
\newcommand{\bn}{{\boldsymbol{n}}}

\newcommand{\bv}{{\boldsymbol{v}}}

\newcommand{\bxi}{\boldsymbol{\xi}}

\newcommand{\rhoper}{{\rho_p}}

\def\pa{\, \cdotp}
\def\vb{\, ,}
\def\va{\,\raise 2pt\hbox{,}}

\def\cF{\mathcal{F}}

\def\cA{\mathcal{A}}

\newcommand{\p}{\partial}

\begin{document}
\title{On the Interplay between Behavioral Dynamics and Social Interactions in Human Crowds}



\maketitle

\centerline{\scshape Nicola Bellomo$^{(1)}$, Livio Gibelli$^{(2)}$, and  Nisrine Outada$^{(3,4)}$}
\medskip
{\footnotesize
	\centerline{$^{(1)}$Department of Mathematics, Faculty of Sciences}
	\centerline{King Abdulaziz University, Jeddah, Saudi Arabia}
}

\medskip
{\footnotesize
	\centerline{$^{(2)}$School of Engineering}
	\centerline{University of Warwick, United Kingdom}
}

\medskip
{\footnotesize
	\centerline{$^{(3)}$Mathematics and Population Dynamics Laboratory-UMMISCO}
	\centerline{Faculty of Sciences of Semlalia of  Marrakech, Cadi Ayyad Univ., Morocco}
}

{\footnotesize
	\centerline{$^{(4)}$Jacques Louis-Lions Laboratory}
	\centerline{Pierre et Marie Curie University, Paris 6, France}
}

\bigskip


\begin{abstract}
This paper provides an overview and critical analysis on the modeling and applications of the dynamics of human crowds, where social interactions can have an important influence on the behavioral dynamics of the crowd  viewed as a living, hence complex, system. The analysis looks at real physical situations where safety problems might arise in some specific circumstances. The approach is based on the methods of the kinetic theory of active particles. Computational applications enlighten the role of human behaviors.
\end{abstract}
\thispagestyle{fancy}
\lfoot{\hspace{0.5cm}\footnotesize \textit{2010 Mathematics Subject Classification} Primary: 82D99, 91A15; Secondary: 91D10. \\ \hspace{0.5cm}\textit{Key words and phrases.} Crowd dynamics, living systems, complexity, kinetic theory, social
dynamics, stress propagation.}


\section{Introduction}

The modeling, qualitative and computational analysis of human crowds is an interdisciplinary research field which involves a variety of challenging analytic and numerical problems,  generated  by the derivation of models as well as by their application  to real world dynamics.

The growing interest for this research field is motivated by the potential benefits for the society. As an example, the realistic modeling of human crowds can lead to simulation tools to support crisis managers to handle emergency situations, as sudden and rapid evacuation through complex venues, where stress induced by overcrowding, or even social conflicts may affect safety of the people~\cite{[EPS02],[KIN13],[LL15],[RNC16],[WCM16]}.

The existing literature on general topics of mathematical modeling of human crowds is reported in some survey papers, which offer to applied mathematicians different view points and modeling strategies in a field, where a unified, commonly shared, approach does not exists yet. More in detail, the review by Helbing~\cite{[HEL01]} presents and critically analyzes  the main features of the physics of crowd viewed as a multi particle system and focuses on the modeling at the microscopic scale for pedestrians undergoing individual based interactions. The survey by Huges~\cite{[HUG03]} and the book~\cite{[CPT14]} deal with the modeling at the macroscopic scale, by methods analogous to those of hydrodynamics, where one of the most challenging conceptual difficulties consists in understanding how the crowd, viewed as a continuum, selects the velocity direction and the speed by which pedestrians move. Papers~\cite{[BBK13],[BG15]} have proposed the concept of the crowds as a living, hence complex, system. This approach requires the search  of mathematical tools suitable to take into account, as far as it is possible, the complexity features of the system under consideration. Scaling problems and mathematical aspects are treated in the book~\cite{[CPT14]}, while the support of modeling to crisis management during evacuation is critically analyzed in the survey~\cite{[BCG16]}.

A critical analysis of the state of the art indicates that the following issue has not yet been exhaustively treated:
 \vskip.1cm
 \textit{The greatest part of known models are based on the assumption of rational, say optimal, behaviors of individuals. However, real conditions can show a presence of irrational behaviors  that can generate events where safety conditions are damaged. When these conditions appear,  small deviations in the input create large deviations in the output. Some of these extreme event are not easily predictable, however a rational interpretation can sometimes explain them once they have appeared. The use of  the term  ``black swan'' a metaphoric expression used by Taleb~\cite{[TAL07]} to denote these events. Derivation of models, and their subsequent validation, should show the ability to reproduce also these extreme events. }
 \vskip.1cm

 Some of the topics mentioned in the above statement have been put in evidence in the review~\cite{[WCM16]}, where it is stressed that modeling approaches should be based on a careful  understanding of human behaviors and that  the majority of current crowd models do not yet  effectively support managers in extreme crisis situations.

Chasing this challenging objective requires acknowledging that the modeling approach can be developed at the three usual scales, namely \textit{microscopic, macroscopic}, and \textit{mesoscopic}, the latter is occasionally called \textit{kinetic}. However none of the aforesaid scaling approaches is fully satisfactory. In fact,  accounting for multiple interactions and for the heterogeneous behavior of the crowd that it empirically observed is not immediate in the case of various known models at the microscopic scale.  This drawback is also delivered by  macroscopic models which kill the aforementioned heterogeneity.

Kinetic type models appear to be more flexible as they can tackle, at least partially, the previously mentioned drawbacks, but additional work is needed to develop them toward the challenging objectives treated in this paper. Namely a multiscale approach is required, where the dynamics at the large scale needs to be properly related to the social dynamics  which appears at the microscopic scale. Some introductory concepts have been proposed in the literature starting from~\cite{[BB15],[BBK13]}, where speed is related to an internal variable of a kinetic model suitable to describe stress conditions by panic.

More recently, \cite{[WSB17]} considers a dynamics  in one space dimension described at the macroscopic scale, where panic is propagated by a BGK type~\cite{[CIP93]} model,  and the velocity is related to panic. All above reasonings indicate that this research topic needs  new ideas  focused on the concept that a crowd is a living system. A study on the role of social dynamics on individual interactions with influence at the higher scale is developed in~\cite{[DAM13],[DLM17]}. Hence, human behaviors have to be taken into account  in the modeling approach.

  Although the literature in the field is rapidly growing and it is already vast, far less developed are the contributions related to the issues that have been enlightened above. Two recent essays can contribute to deal with the aforementioned objective. In more detail, the recent paper~\cite{[ABG16]} has proposed  a new system sociology approach to the modeling a variety  of social phenomena, while learning dynamics in large populations is dealt with in~\cite{[BDG16],[BDG16B]}. This papers develop an approach based on kinetic theory methods for active particles, namely by methods that show some analogy with the mesoscopic approach to crowd modeling.

  Our paper is devoted to modeling the complex interaction between social and mechanical dynamics. The paper also accounts for  the additional difficulty of the modeling of the quality and geometry of the venues where the dynamics occurs and the interaction of walkers  with obstacles and walls. Still the role of the quality of the venue, which is an important feature that modifies the speed of walkers in a crowd, is treated in our paper. In more detail, the contents of this paper is presented as follows:

Section 2  defines the class of social phenomena that our paper aims at including in the modeling approach. Subsequently, the selection of the mesoscopic scale is motivated in view of a detailed analysis  to be developed in the following sections. This conceptual background is presented toward the strategic objective of designing mathematical models suitable to depict the complexity features of a social crowd.

Section 3 deals with the modeling for a crowd of individuals belonging to different groups, where a common different way of organizing the dynamics and the interactions with other individuals is shared. This section transfers into a mathematical framework the general concepts, presented in Section 2,  with the aim of providing the conceptual basis of the derivation of models that can be obtained by inserting into this structure models suitable to describe interactions at the micro-scale. This section also enlightens the improvements of our paper with respect to the existing literature.

Section 4 shows how two specific models can be derived according to the aforementioned general structure. The first model describes the onset and propagation of panic in a crowd starting from a localized onset of stress conditions. The second model includes the presence of leaders who play the role of driving the crowd out of a venue in conditions where panic propagates.

Section 5 presents some simulations which provide a pictorial description of the dynamics. Suitable developments of Monte Carlo particle methods, starting from~\cite{[ARI01],[BFG15],[BIR94],[PT14]}, are used. Simulations enlighten specific features of the patterns of the flow focusing specifically on the evacuation time and the concentration high density that can induce incidents.

Section 6 presents a critical analysis of the contents of the paper as well as  an overview of research perspectives which are mainly focused on multiscale problems.

\section{Complexity Features of Social Crowds}

This section presents a phenomenological description of the social and mechanical features which should be taken into account in modeling of social crowds. The various models proposed in the last decades were derived referring to a general mathematical structure, suitable to capture the complexity features of large systems of interacting entities, and hence suitable to provide the conceptual basis towards basis for the derivation of specific models, which are derived by implementing  the said structure by heuristic models  of individual based interactions. On the other hand, recent papers by researchers involved in the practical management of real crowd dynamics problems, including crisis and safety problems, have enlighten that social phenomena pervade heterogeneous crowds and can have an important influence on the interaction rules~\cite{[BCG16],[CMV15],[HFV00],[HJ09],[HJA07],[MHG09],[MT11],[RNC16],[WCM16],[WJJ13],[WIN12]}. Therefore, both social and mechanical dynamics, as well as their complex interactions, should be taken into account.
A kinetic theory approach to the modeling of crowd dynamics in the presence of social phenomena, which can modify the rules of mechanical interactions, has been proposed in our paper. Two types of social dynamics have been specifically studied, namely the propagation of stress conditions and the role of leaders. The case study proposed in Section 5 has shown that stress conditions can induce important modifications in the overall dynamics and on the density patterns thus enhancing formation of overcrowded zones. The specific social dynamics phenomena studied in our paper have been motivated by situations, such as fire incidents or rapid evacuations, where safety problems can arise~\cite{[EPS02],[LL15],[RNC16],[WCM16]}.

The achievements presented in the preceding sections motivate a systematic computational analysis focused on a broader variety of case studies focusing specifically to enlarge the variety of social phenomena inserted in the model. As an example, one might consider even extreme situations, where antagonist groups contrast each other in a crowd. This type of developments can be definitely inserted into a possible research program which is strongly motivated by the security problems of our society.

Furthermore, we wish returning to the scaling problem, rapidly introduced in Sections 1 and 2, to propose a critical, as well as self-critical, analysis induced also by the achievements of our paper on the modeling human behaviors in crowds. In more detail, we observe that it would be useful introducing aspects of social behaviors also in the modeling at the microscopic and macroscopic scale. Afterwards, a critical analysis can be developed  to enlighten advantages and withdraws of the selection of a certain scale with respect to the others.

This type of analysis should not hide the conceptual link which joins the modeling approach at the different scales. In fact a detailed analysis of individual based interactions (microscopic scale)  should implement the derivation of kinetic type models (mesoscopic scale), while hydrodynamic models (macroscopic scale) should be derived from the underlying description delivered kinetic type models by asymptotic methods where a small parameter corresponding to the distances between individuals is let to tend to zero.  Often models are derived independently at each scale, which prevents a real multiscale approach.

Some achievements have already been obtained on the derivation of macroscopic equations from the kinetic type description for crowds in unbounded domains~\cite{[BB15]} by an approach which has some analogy with that developed for vehicular traffic~\cite{[BBNS14]}. However, applied mathematicians might still investigate how the structure of macroscopic models is modified by social behaviors. This challenging topic might be addressed even to the relatively simpler problem of vehicular traffic where individual behaviors are taken into account~\cite{[BDF17]}.

 Finally, let us state that the ``important'' objective, according to our own bias, is the development of a systems approach to crowd dynamics, where models derived at the three different scales might coexist in complex venues where the local number density from rarefied to high number density. This objective induces the derivation of models at the microscopic scale consistent with models at the macroscopic scale with the intermediate description offered by the kinetic theory approach.

We do not naively claim that models can rapidly include the whole variety of social phenomena. Therefore, this section proposes a modeling strategy, where only  a number of them is selected. An important aspect of the strategy is the choice of the representation and modeling scale selected referring to the classical scales, namely microscopic (individual based), macroscopic (hydrodynamical), and mesoscopic (kinetic). The sequential steps of the strategy are as follows:

\begin{enumerate}

\vskip.1truecm  \item  Assessment of the complexity features of crowds viewed as living systems;

\vskip.1truecm  \item  Selection of the social phenomena to be inserted in the model;

\vskip.1truecm  \item Selection of the modeling scale and  derivation of a  mathematical structure consistent with the requirements in the first two items;

\vskip.1truecm  \item Derivation of models by inserting,  into the said structure, the mathematical description of interactions for both social and mechanical dynamics including their reciprocal interplay.

\end{enumerate}

The structure mentioned in Item~3. should be  general enough to include a broad variety of social dynamics. However, the derivation of models mentioned in Item~4. can be effectively specialized only if specific case studies are selected. Some rationale is now proposed, in the next subsections,  for each of these topics referring to the existing literature so that repetitions are avoided.

\subsection{Complexity features} The recent literature on crowd modeling~\cite{[BBK13],[BG15],[BG16]} has enlightened the need of modeling crowd dynamics, where the behavioral features of crowds to be viewed as a living, hence complex system, are  taken into account. Indeed, different behaviors induce different interactions and hence walkers' trajectories. The most important feature is the ability to express a strategy which is heterogeneously distributed among  walkers and depends on their own state and on that of the entities in their surrounding walkers and environment. Heterogeneity can include a possible presence of leaders, who aim at driving the  crowd to their own strategy. As an example,  leaders can contribute, in evacuation dynamics, to drive walkers toward appropriate strategies including the selection of optimal routes among the available ones.

\subsection{Selection of social phenomena} The importance of understanding human behaviors in crowds is undisputed~\cite{[RNC16],[WCM16]} as they can have an important influence on the individual and collective dynamics and can contribute to understand crisis situations and support their management~\cite{[BCG16]}.

A crowd might be subdivided into different groups due both to social and mechanical features which have to be precisely referred to the type of dynamics which is object of modeling. Examples include the presence of leaders as well as of  stress conditions which, in some cases, are induced by overcrowding. In some cases, a crowd in a public demonstration includes the presence of  groups of rioters, whose aim is not the expression of a political-social opinion, but instead to create conflict with security forces. These examples should be made more precise when specific case studies are examined.

An important topic, is the role of irrational behaviors, where these emotional states can be induced by perception of danger~\cite{[FPA11]} or simply by overcrowding. Our interest consists in understanding which type of collective behavior develop in different social situations and how this behavior propagates. Indeed, we look at a crowd in a broad context, where different social phenomena can appear~\cite{[ABG16]}.

\subsection{Modeling interactions} Interactions are nonlinearly additive and  refer both to mechanical and social dynamics and include the way by which walkers adjust their dynamics to  the specific features of the venue, where they move. Propagation of social behaviors has to be modeled as related to interactions.

A key example is given by the onset and propagation of stress conditions, which may be generated in a certain restricted area and then diffused over the whole crowd. These conditions can have an important influence over dynamical behaviors of walkers \cite{[HJA07]}.

The so called \textit{faster-is-slower} effect, namely increase of the individual speed but toward congested area, rather than the optimal directions, which corresponds to an increase of evacuation time in crisis situation that require exit from a venue. In addition, stress conditions can break cooperative behaviors inducing irrational selfishness. This topic was introduced in~\cite{[BBK13]}, where it was shown how an internal variable can be introduced to model stress conditions which modify flow patterns, see also~\cite{[BG15],[WSB17]}.

Recent research activity on empirical data has been addressed to acquire information on the estimate of forces exchanged by pedestrians~\cite{[CMV15]} and on crowd behaviors including aggregation phenomena by coarse grain observation. These investigations go beyond the study of velocity diagrams~\cite{[MMGM17]}.

\subsection{Scaling and derivation of mathematical structures:} The mesoscopic description is based on  kinetic theory methods, where the representation of the system is delivered by a suitable probability distribution over the microscopic state of walkers, which is still identified by  the  individual  position and  velocity,  however additional parameters  can be added such as size, and variables to model the social state. Models describe the dynamics of this distribution function by  nonlinear integral-differential equations. As it is known, none of the aforesaid scaling approaches, namely at the microscopic, macroscopic, and mesoscopic scales, are fully satisfactory. In fact, known models at the microscopic scale do not account for multiple interactions and it may difficult, if not impossible, to use data from microscopic observations to infer the crowd dynamics in a different but similar situation. On the other hand, the heterogeneous behavior of pedestrians get lost in the averaging process needed to derive the macroscopic models which therefore totally disregard this important feature.
Mesoscale (kinetic) models appear to be more flexible as they can tackle the previously mentioned drawbacks, but additional work is needed to develop them toward the challenging objectives treated in this paper.

\subsection{Derivation of models:} The  derivation requires the modeling of interactions among walkers and the insertion of these models into a mathematical structure consistent with the aforementioned scale as well as with the specific phenomenology of the system under consideration.

 This approach needs a deep understanding of the psychology and emotional states of the crowd. The modeling approach should  depict how the heterogeneous distribution evolves in time.  The conceptual difficulty consists in understanding how emotional states can modify the rules of interactions.  Therefore, the modeling approach should include all of them, heterogeneity, as well as the heterogeneous behavior of individuals and the growth of some of them also induced by collective learning~\cite{[BDG16]}.

Furthermore, the features of the venue where walkers move cannot be neglected, as enlightened in~\cite{[RGP13],[RRP16]}, as it can have an influence on the speed due both to mechanical actions, for instance the presence of stairs, or to emotional states which can induce aggregation or disaggregation dynamics.
All dynamics need to be properly referred to the geometrical and physical features of the venue which, at least in principles, might be designed according to well defined safety requirements~\cite{[RON15],[RKN16]}.

\section{On a Kinetic Mathematical Theory of  Social Crowd Dynamics}

 This section deals with the derivation of models by suitable developments of  the kinetic theory for active particles~\cite{[BKS13]}. Our approach focuses on heterogeneous human crowds in domains with boundaries, obstacles and walls. According to this  theory, walkers are considered \textit{active particles}, for short a-particles, whose state is identified, in addition to mechanical variables, typically position and velocity, by an additional variable modeling their emotional or social state called  \textit{activity}. These particles can be subdivided into \textit{functional subsystems}, for short FSs, grouping a-particles that share the same \textit{activity} and mechanical purposes, although if heterogeneously within each  FS.

 The theoretical approach to modeling aims  at transferring into a formalized framework the phenomenological description proposed in Section 2. This objective can be achieved in the following sequential steps:

\begin{enumerate}

\item Assessment of the possible dynamics, mechanical and social, which are selected toward the modeling approach, and representation of social crowds;

\vskip.1cm \item  Modeling  interactions;

\vskip.1cm \item Derivation of  a mathematical structure  suitable to provide the conceptual basis for the derivation of specific models.

\end{enumerate}

These sequential steps are treated in the following subsections. The modeling approach proposed in our paper includes a broad variety of mechanical-social dynamics that have not  yet been treated exhaustively in the literature.

\subsection{Mechanical-social dynamics and representation}

Let us now provide a detailed description of the specific features that our paper takes into account in the search for a mathematical structure accounted for deriving mathematical models of social crowds.

\begin{itemize}

\item The a-particles are heterogeneously distributed in the crowd which is subdivided into groups labeled by the subscript $i=1, \ldots, n$,  corresponding to different functional subsystems.

    \vskip.1cm \item The mechanical state of the a-particles  is defined by position $\bx$,  velocity $\bv$, while their emotional state modeled by a  variable at the microscopic scale, namely the  activity, which takes value in the domain $[0,1]$ such that $u=0$ denotes the null expression,  while $u=1$ the highest one.

\vskip.1cm \item If the overall crowd moves toward different walking directions a further subdivision can be necessary to account for them.

\vskip.1cm \item Interactions lead not only to modification of mechanical variables, but also of the activity which, in turn, modifies the rules of mechanical interactions.

\end{itemize}

According to this description, the  \textit{microscopic state} of the a-particles, is defined by position $\bx$,  velocity $\bv$,  and activity  $u$.  Dynamics in two space dimensions is considered, while polar coordinates are used for the velocity variable, namely $\bv = \{v, \theta\}$, where $v$ is the speed and $\theta$ denotes the velocity direction. Dimensionless, or normalized, quantities are used  by referring the components of $\bx$ to a characteristic length $\ell$, while the velocity modulus is divided by the limit velocity, $V_\ell$, which can be reached by a fast pedestrian in free flow conditions; $t$ is the dimensionless  time variable obtained referring the  real time to a suitable critical time $T_c$ identified by the ratio between $\ell$ and $V_\ell$. The limit velocity depends on the quality of the environment, such as presence of positive or negative slopes, lighting and so on.

The  \textit{mesoscopic (kinetic) representation} of  each FS  is delivered  by the statistical distribution at time $t$, over the microscopic state:
\begin{equation}
	f_i=f_{i}(t,\,\bx,\,v, \theta, u), \quad \bx \in \Sigma \ \subset \mathbb{R}^3, \quad v\in [0,1], \quad \theta \in [0,2\pi) \quad u \in [0,1].
\end{equation}

If $f_{i}$ is locally integrable then $f_{i}(t,\,\bx,\,\bv, u)\,d\bx\,d\bv\,du$ is the (expected) infinitesimal number of pedestrians  of the i-th FS whose micro-state, at time $t$, is comprised in the elementary volume $[\bx , \bx + d\bx] \times [\bv , \bv + d\bv] \times [u , u + du]$ of the space of the micro-states, corresponding to the variables space, velocity and activity. The statistical distributions $f_{i}$ are divided by $n_M$, which defines the maximal full packing density of pedestrians and it is assumed to be approximately seven walkers per square meter.

\textit{Macroscopic observable quantities} can be obtained, under suitable integrability assumptions, by weighted  moments of the distribution functions. As an example, the local \textit{density} and \textit{mean velocity} for each  $i$-FS reads
\begin{equation}\label{mac-1}
\rho_i (t, \bx) = \int_0^1 \int_0^{2\pi}  \int_0^1 f_{i}(t,\,\bx,\,v, \theta, u)\,v dv\, d\theta\,du,
\end{equation}
and
\begin{equation}\label{mac-2}
\boldsymbol{\xi}_i (t, \bx) = \frac{1}{\rho_i (t,\bx)} \int_0^1 \int_{0}^{2\pi} \int_0^1   \bv\, f_{i}(t,\,\bx,\,v, \theta, u)\,v dv\, d\theta\,du,
\end{equation}
whereas global expressions are obtained by summing over all $i$ indexes
\begin{equation}\label{mac-3}
\rho (t,\bx) = \sum_{i=1}^n \rho_{i} (t,\bx), \hspace{1cm} \mbox{and} \hspace{1cm} \boldsymbol{\xi} (t,\bx) = \frac{1}{\rho (t, \bx)} \, \sum_{i=1}^n \,\rho_i (t,\bx) \boldsymbol{\xi}_{i} (t, \bx).
\end{equation}

Specific applications might require computation of marginal densities such as the local mechanical distribution and the local activity distribution in each FS:
\begin{equation}
f_{i}^M (t, \bx, \bv) = \int_0^1 f_{i}(t,\,\bx,\,v, \theta, u)\,du,
\end{equation}
and
\begin{equation}
f_{i}^A (t, \bx, u) = \int_0^1 \int_{0}^{2\pi}   f_{i}(t,\,\bx,\,v, \theta, u)\,v dv\, d\theta.
\end{equation}

\subsection{Modeling interactions}

Interactions correspond to  a decision process by which each active particle modifies its activity and decides its mechanical dynamics depending on the micro-state and distribution function of the neighboring  particles in its  interaction domain. This process modifies  velocity direction and speed.  Interactions involve,  at each time $t$ and for each FS,  three types of a-particles: The \textit{test particle}, the \textit{field particle}, and the \textit{candidate particle}. Their distribution functions are, respectively $f_i(t,\bx, \bv, u)$,  $f_k(t, \bx,  \bv^*, u^*)$, and  $f_h(t, \bx,  \bv_*, u_*)$. The test particle, is representative, for each FS, of the whole system, while the candidate particle can acquire, in probability, the micro-state of the test particle after interaction with the field particles. The test particle loses its state by  interaction with the field particles.

Interactions  can be modeled using the following quantities: \textit{Interaction domain} $\Omega_s$,
\textit{interaction rate} $\eta$,   \textit{transition probability density} $\cA$, and the overall action of the field particles. These quantities can depend on the micro-state and on the distribution function of the interacting particles, as well as on the quality of the venue-environment where the crowd moves. The  definition of these terms, is reported in the following, where the terms $i$-particle  is occasionally used to denote  a-particles  belonging to the $i$-th FS.

\begin{itemize}

\item  \textit{Short range interaction domain:} A-particles interact with the other a-particles in a domain $\Omega_s$  which is a circular sector, with radius $R_s$, symmetric with respect to the velocity direction being defined by the visibility angles $\Theta$ and $- \Theta$. The a-particles perceives in $\Omega_s$ local  density and density gradients.

\vskip.1cm \item   \textit{Perceived density:} Particles moving along the direction $\theta$ perceive a density $\rho^p_\theta$ different from the local density $\rho$. Models should account that $\rho^p_\theta > \rho$ when the density increases along $\theta$, while $\rho^p_\theta < \rho$, when the density decreases.

\vskip.1cm \item  \textit{Quality of the venue} is a local quantity modeled by the parameter $\alpha = \alpha (\bx) \in [0,1]$, where $\alpha = 0$ corresponds the worse conditions which prevent motion, while $\alpha = 1$ corresponds to the best ones, which allows a rapid motion.

\vskip.1cm \item  \textit{Interaction rate}  models the frequency by which a candidate (or test)  $h$-particle in $\bx$ develops contacts, in $\Omega_s$, with a field $k$-particle. The following notation is used $\eta_{hk}[\f](\bx,\bv_*,\bv^*, u_*, u^*; \alpha)$.

\vskip.1cm \item  \textit{Transition probability density:} $\cA_{hk}^i[\f](\bv_* \to \bv, u_* \to u|\bv_*,  \bv^*, u_*, u^*; \alpha)$
models the probability density that a candidate $h$-particle in $\bx$ with state $\{\bv_*, u_*\}$ shifts to the state of the $i$-test particle due to the interaction with  a field $k$-particle in $\Omega_s$ with state  $\{\bv^*, u^*\}$.

\vskip.1cm \item  \textit{The overall action of the field particles:} $\cF(t, x,\bv, u, \alpha)$ which describes the average action of the field particles, in the interaction domain $\Omega_s$, over the test particle with the state $\{\bx, \bv, u\}$, and is defined by
\begin{equation}
\label{eq:transport1}
\cF(t, \bx,\bv, u, \alpha) = \ve \int_{\Omega_s} \, \vf(\bx,\bx_*, \bv, \bv_*,u, u_*,\alpha)f(t,\bx_*,\bv_*,u_*)\,d\bx_*\,d\bv_*\,du_*,
\end{equation}
where $\vf(\bx,\bx_*, \bv, \bv_*, u, u_*,\alpha)$ models the action at the microscopic scale between the field and the test particle and $\varepsilon$ corresponds to the scaling related to the independent variables.
\end{itemize}

\subsection{Derivation of a mathematical structure}

Let us now consider the derivation of a general structure suitable to include all types of interactions presented in the preceding subsection.  This approach aims at overcoming the lack of first principles that govern the living matter. Indeed, such structure claims to be consistent with the complexity features of living systems~\cite{[BKS13]}. The  mathematical structure consists in an integro-differential equation suitable to describe the time dynamics of the distribution functions $f_i$. It  can be obtained by a balance of particles in the elementary volume, $[\bx, \bx + d\bx] \times [\bv, \bv + d\bv] \times [u, u + du]$, of the space of the micro-states. This conservation equation corresponds to equating  the variation rate of the number of active particles plus the transport due to the velocity variable and the acceleration term to net  flux rates within the same FS and  across FSs.

It is worth stressing that this structure is consistent with the paradigms presented in Section 2. In more detail, ability of pedestrians to express walking strategies based on interactions with other individuals is modeled by the transition probability density, while  the heterogeneous distribution of the said  strategy (behavior) corresponding both to different  psycho-logic attitudes and mobility abilities is taken into account by the use of a probability distribution over the  mechanical and activity variables.  Interactions have been assumed to be nonlocal and nonlinearly additive as the strategy developed by a pedestrian is a nonlinear combination of different stimuli generated by interactions with other pedestrians and with the external environment.

We consider different structures, which progressively account for dynamics that include a reacher and reacher dynamics. All of them  are obtained by a balance of number of particles in the elementary volume of the space of microscopic states.

\vspace{.2cm} \noindent $\bullet$ \textbf{One component crowd:}  In the case of only one FS,
namely  $n=1$,  the subscripts can be dropped and crossing FSs is not included. Hence, the balance of particles yields:
\begin{eqnarray}
\label{StructureI}
&&\left(\p_t  +  \bv \cdot \p_\bx \right)\, f(t, \bx, \bv, u) = A[\f](t, \bx, \bv, u)  \nonumber  \\
&& \hskip.5truecm =  \int_{D \times D} \eta[f](\bx,\bv_*,\bv^*,u_*, u^*;\alpha)
\, \cA[f](\bv_* \to \bv, u_* \to u|\bv_*,  \bv^*, u_*, u^*; \alpha) \nonumber \\
 && \hskip2truecm \times  \, f(t, \bx, \bv_*, u_*) f(t, \bx,  \bv^*, u^*)\,d\bv_*\,d\bv^* \, du_*\, du^* \vb \nonumber \\
 && \hskip1truecm - f(t, \bx, \bv,u) \int_{D} \eta[f](\bx,\bv,\bv^*,u, u^*;\alpha) \,f(t, \bx,  \bv^*, u^*)\, d\bv^* \, du^*,
 \end{eqnarray}
where $D = [0,1] \times [0, 2\pi) \times [0,1]$.

\vspace{.2cm} \noindent $\bullet$ \textbf{Multicomponent crowd without FS-crossing:} Corresponding to the case of multiple FSs while the dynamic across them is not included. The mathematical structure in this case, using the simplified notation $\cA_{ik}:= \cA_{ik}^i$, reads:
\begin{eqnarray}
\label{StructureII}
&&\left(\p_t  +  \bv \cdot \p_\bx \right)\, f_i(t, \bx, \bv, u) = P_i[\f](t, \bx, \bv, u) \nonumber  \\
&& \hskip.5truecm = \sum_{k=1}^n \,  \int_{D^2} \eta_{ik}[\f](\bx,\bv_*,\bv^*, u_*, u^*;\alpha) \, \cA_{ik}[\f](\bv_* \to \bv, u_* \to u|\bv_*,  \bv^*, u_*, u^*; \alpha) \nonumber \\
 && \hskip2truecm \times \, f_i(t, \bx, \bv_*, u_*) f_k(t, \bx,  \bv^*, u^*)\,d\bv_*\,d\bv^* \, du_*\, du^*, \nonumber \\
 && \hskip1truecm -  f_{i}(t, \bx, \bv,u)  \sum_{k=1}^n \int_{D} \eta_{ik}[\f](\bx,\bv,\bv^*,u, u^*;\alpha) \, f_{k}(t, \bx,  \bv^*, u^*)\, d\bv^* \, du^*.
\end{eqnarray}

\vspace{.2cm} \noindent $\bullet$ \textbf{Multi-component crowd with FSs crossing.} Corresponding to the general case of multiple FSs and where the dynamic across them is taken into account:
\begin{eqnarray}
\label{StructureIII}
&&\left(\p_t  +  \bv \cdot \p_\bx \right)\, f_i(t, \bx, \bv, u) = Q_i[\f](t, \bx, \bv, u) \nonumber  \\
&&\hskip.5truecm  = \sum_{h=1}^n  \sum_{k=1}^n  \int_{D \times D} \eta_{hk}[\f](\bx,\bv_*,\bv^*, u_*,u^*;\alpha)\nonumber  \\
&&\hskip2truecm  \times \cA_{hk}^i[\f](\bv_* \to \bv, u_* \to u| \bv_*,  \bv^*, u_*, u^*;\alpha) \nonumber\\
 && \hskip2truecm \times  \, f_{h}(t, \bx,\bv_*,u_*) f_{k}(t, \bx, \bv^*, u^*) \, d\bv_* \, d\bv^* \, du_* \, du^*\nonumber \\
   && \hskip.5truecm - f_{i}(t, \bx, \bv, u) \sum_{k=1}^n \int_{D} \eta_{ik}[\f](\bx,\bv,\bv^*, u, u^*;\alpha)f_{k}(t, \bx, \bv^*, u^*)\, d\bv^*\, du^*.
 \end{eqnarray}

\vspace{.2cm} \noindent $\bullet$ \textbf{One component crowd with long large interactions.} In large venues, aggregation of walkers can also be modeled by long range interactions involve test particles interacting with field particles. Interactions occur in a visibility domain $\Omega_v$ which can be defined as a circular sector, with radius $R_v$, symmetric with respect to the velocity direction being defined by the visibility angles $\Theta$ and $- \Theta$. These  interactions can modify the activity variable depending on the distance of the interacting pairs. The mathematical structure \eqref{StructureI} is then modified as follows:
\begin{small}
\begin{equation}
\label{StructureIV}
\left(\p_t  +  \bv \cdot \p_\bx \right)\, f(t, \bx, \bv, u) +  \cT(f)(t, \bx,\bv, u) = A[\f](t, \bx, \bv, u),
\end{equation}
\end{small}
where  the acceleration term $\cT$ is defined by
\begin{equation}
\label{eq:transport2}
\cT(f)(t, \bx,\bv,u) = \varepsilon \, \partial_v \big(\cF(t, \bx,\bv,u,\alpha)\, f(t, \bx,\bv,u) \big),
\end{equation}
while the  term $A$ modeling the net flux and can be  formally defined by  Eq.~\eqref{StructureII}.

In \eqref{StructureI}-\eqref{StructureIV} round and square parenthesis distinguish, respectively, the argument of linear and nonlinear interactions, where linearity involves only microscopic and independent variables, while nonlinearity involves the dependent variables, namely the distribution function and/or its moments. In addition,  these terms are nonlocal and depend on the quality of the environment-venue. The derivation of specific models can clarify this matter.

\section{From the  Mathematical Structure to Models}

The mathematical structures presented in the preceding section provide the conceptual framework for the derivation of models which can be obtained by  selecting the functional subsystems relevant to the specific study to be developed and by  modeling interactions related to the strategies developed by a-particles within each subsystem. This section shows how  certain models of interest for the applications, selected among various possible ones, can be derived. Subsequently, Section 5  investigates, by appropriate simulations, their predictive ability.

 In more detail, we look for models suitable to understand how the stress propagates in the crowd and how the flow patterns are subsequently modified with respect to the initial flow conditions. An additional topic consists in understanding how  the flow patterns can be modified by the presence of leaders. The derivation of  models is proposed in the next two subsections, the first of the two deals with the modeling of the crowd in absence of leaders, while the second subsection shows how the modeling approach can account for the presence of leaders.

\subsection{Dynamics with stress propagation}

Let us consider a crowd in a venue of the type represented in Fig.~1.  Only one FS is considered, therefore the mathematical structure used towards the modeling is given by Eq.~(\ref{StructureII}) for a system whose state is described by the distribution function $f = f(t, \bx, \bv, u)$. Let us consider the modeling of the various terms of the said structure.

\vskip.2cm \noindent \textit{Modeling the limit velocity:} The limit velocity depends on the quality of the venue. A simple assumption is as follows: $V_\ell = \alpha V_L$, where $V_L$ is the limit velocity in an optimal environment.

\vskip.2cm  \noindent \textit{Modeling the encounter rate:}  A simple assumption  consists in supposing that  that it grows with the activity variable and with the perceived density starting from a minimal value $\eta_0$, namely
\begin{equation}
\eta = \eta[f] = \eta_0(1 + \beta\, u\, \rho^\theta [f]),
\end{equation}
where $\beta$ is a positive defined constant. A minimal model is obtained with $\eta \cong \eta_0$.

\vskip.2cm \noindent \textit{Modeling the dynamics of interactions:} These interactions correspond to a \textit{decision process} by which, following the rationale of~\cite{[BG15]}, each walker develops a strategy obtained by the following sequence of decisions: (1) Exchange of the emotional state; (2) Selection of the walking direction; (3) Selection of the walking speed. Decisions are  supposed to be sequentially dependent and to occur with an encounter rate related to the local flow conditions. Hence, the process corresponds to the following factorization:
\begin{equation}\label{factorization}
 \cA(\bv_* \to \bv, u^* \to u) = \cA^u(u^* \to u)\times \cA^\theta (\theta_* \to \theta) \times \cA^v(v_* \to v).
\end{equation}

Starting from this assumption, a simple model can be obtained for each of the three types of dynamics under the additional assumption that the output of the interaction is a delta function over the most probable state:

\vskip.1cm \noindent \textit{1. Dynamics of the emotional state:} The dynamics by which the stress initially in $\Sigma_s$ diffuse among all walkers is driven by the highest value, namely:
\begin{equation}\label{Au-1}
u^* > u_* \, : \quad \mathcal{A}^u(u_* \to u|u_*,u^*)= \delta \big(u - \ve(u^* - u_*)(1 - u_*) \big),
\end{equation}
and
\begin{equation}\label{Au-2}
u^* \leq u_* \, : \quad \mathcal{A}^u(u_* \to u|u_*,u^*)= \delta \big(u - u_* \big).
\end{equation}

\vskip.1cm \noindent  \textit{2. Dynamics of the velocity direction:} It is expected that at high density, walkers try to drift apart from the more congested area moving in the direction of $\bnu_V$ (direction of the less congested area), while  at low density, walkers head for the target identified  $\bnu_T$ (the exit door) unless their level of anxiety is high in which case they tend to follow the mean stream as given by  $\bnu_S$ (direction of the stream).  Walkers select the velocity direction $\theta$ by an individual estimate of the local flow conditions and consequently develop a decision process which leads to the said directions.
The sequential steps of the process are:
\begin{enumerate}
\item Perception of the density $\rho$ which has an influence on the attraction to $\bnu_T$, as it increases by decreasing density.
 \item Selection of a walking direction  between  the attraction to  $\bnu_S$ and the search of less congested areas is identified by the direction given by the unit vector $\bnu_V$, where this selection is based on the assumption that increasing $\beta$  increases the attraction to $\bnu_S$ and decreased that to $\bnu_T$ increases.
 \item Accounting for  for the presence of walls which is modified by the distance from the wall $d_w$ supposing that the search of less congested areas decreases with decreasing distance which induces an attraction toward  $\bnu_T$.
\end{enumerate}

 The selection of the preferred walking direction $\theta$ is in two steps: first the walker in a point $P$ selects a direction  $\theta_1$, then if the new direction effectively moves toward the exit area, then  $\theta_1$ is not modified. On the other hand, if it is directed toward a point $P_w$ of the boundary then the direction is modified by a weighted choice between  $\theta_1$ and the direction from the position $\theta_T$ from $P$ to $T$, where the weight is given by the distance $d_w = |P - P_w|$. Accordingly, the transition probability density for the angles is thus defined as follows:
\begin{equation}\label{transition-direction}
\mathcal{A}_\theta [\rho,\bx] (\theta_* \to \theta) = \delta \left(\theta -\theta_* \right), \quad \hbox{with} \quad \theta = (1 - d_w) \theta_T + d_w \theta_1,
\end{equation}
where $d_w$ is assumed to be equal to one $d_w = 1$ if $\theta_1$ is directed toward $T$, and where $\theta_1$ is given by:
\begin{equation}
\theta_1 [\rho, \bx, u] = \frac{\rho  \bnu_V  + \left(1-\rho \right)
             \displaystyle{\frac{u  \bnu_S + (1-u)\bnu_T}{\left\| u \bnu_S + (1-u) \bnu_T \right\|}}}
            {\left\| \rho \bnu_V + \left(1-\rho \right)
             \displaystyle{\frac{u \bnu_S + (1-u) \bnu_T}{\left\|u \bnu_S + (1-u) \bnu_T \right\|}} \right\|}\va
\end{equation}
where
\begin{equation}
 \bnu_V = -\frac{\nabla_{\bx} \rho}{\left\| \nabla_{\bx} \rho \right\|}\va \hspace{2cm}
 \bnu_S = \frac{\bxi}{\left\| \bxi  \right\|}\pa
\end{equation}

\vskip.2cm \noindent  \textit{3. Perceived density:} Walkers moving along a certain direction perceive a density higher (lower) than the real one in the presence of positive (negative) gradients. The following model has been proposed in~\cite{[BG15]}:
\begin{equation}
\label{eq:rho_perceived}
\rhoper  = \rhoper [\rho]= \rho + \frac{\partial_{p} \rho}{\sqrt{1 + (\partial_{p} \rho)^2}}\,
                           \left[(1 - \rho)\, H(\partial_{p} \rho) + \rho \, H(- \partial_{p} \rho)\right]\va
\end{equation}
where
$\partial_p$ denotes the derivative along the direction $\theta^{(p)}$ while $H(\cdot)$ is the Heaviside function $H(\cdot \geq 0) = 1$,
and $H(\cdot < 0) = 0$. The density $\rhoper$ delivered by this model takes value in the domain $[0,1]$.

\vskip.2cm \noindent  \textit{4. Dynamics of the speed:}  Once the direction of motion has been selected,  the walker adjusts the speed to the local density and mean speed conditions. A specific model, in agreement with~\cite{[BG16]} can be used:

\noindent $\bullet$ If $\xi \geq v_*$:
\begin{equation}
\label{speedI}
 \cA^v(v_* \to v) = p_a(\alpha, u,  \rho) \, \delta(v - \xi_a(\alpha, u,  \rho))
 + (1 -  p_a(\alpha, u,  \rho))\, \delta(v - v_*),
 \end{equation}
 and,  if $\xi > v_*$:
 \begin{equation}
\label{speedII}
\cA^v(v_* \to v) =  p_d(\alpha, u,  \rho) \delta(v -\xi_d(\xi,\rhoper))
 + (1 -  p_d(\alpha, u,  \rho))\, \delta(v - v_*)
 \end{equation}
where
$$
p_a (\alpha, u,  \rho) = \alpha \, u (1 - \rhoper), \qquad \xi_a (\alpha, u,  \rho)=  \xi + \alpha\, u(1 - \rhoper)(\alpha u - \xi),
$$
and
$$
p_d(\alpha, u,  \rho) = (1 -  \alpha \, u) \rhoper, \qquad \xi_d(u,  \rho) =  \xi (1 - \rhoper).
$$

This heuristic model corresponds to the following dynamics: \textit{If the walker's speed is lower than the mean speed, then the model describes a trend of the walker increase the speed by a decision process which is enhanced by low values of the perceived density and by the goodness are the quality of the venue. The opposite trend is modeled when the walker's speed is lower than the mean speed.}

This model which is valid if  $\alpha \, u < 1$ has shown to reproduce realistic velocity diagrams, where the mean velocity decays with the density by a slope which is close to zero for $\rho =0$ and $\rho =1$. In addition, the diagram decreases when the quality of the venue and the level of anxiety decreases~\cite{[BG16]}. Of course, it is a heuristic model based on a phenomenological interpretation of reality. Therefore it might be technically improved.

\subsection{Modeling the presence of leaders}

This section develops a model where a number of leaders are mixed within the crowd. The aim of the modeling consists in understanding how their presence modifies the dynamics. Two FSs are needed to represents the overall systems, while additional work on modeling interactions has to be developed. In consonance with to the modeling approach proposed in the
preceding section, the following subdivision is proposed: $i=1$ \textit{walkers }, $i=2$ \textit{leaders }. The main features of the interactions that a candidate (or test) particle can undergo is sketched in the following;
$$
\begin{array}{ll}
\mbox{$\bullet$ Within the same FS} & \left\lbrace
\begin{array}{l}
\mbox{(i) Interactions between a \textit{walker}} \\
\mbox{\hspace{.4cm} and the field \textit{walkers} (\textit{I-WW}).}\vspace{0.2cm}\\
\mbox{(ii) Interactions between a \textit{leader}} \\
\mbox{\hspace{.6cm} and the field \textit{leaders} (\textit{I-LL}).}
\end{array}\right.
\\
{}
\\
{}
\\
\mbox{$\bullet$ Within different FSs} & \left\lbrace
\begin{array}{l}
\mbox{(iii) Interactions between a \textit{walker}} \\
\mbox{\hspace{.6cm} and the field \textit{leaders} (\textit{I-WL}).}\vspace{0.2cm}\\
\mbox{(iv) Interactions between a \textit{leader}} \\
\mbox{\hspace{.6cm} and the field \textit{walkers} (\textit{I-LW}).}
\end{array}\right.
\\
\end{array}
$$

The representation of the system is delivered by the normalized probability distributions
\begin{equation}
f_1, \, f_2: \quad  [0,T[ \times \Sigma \times [0,1] \times [0,2\pi) \times [0,1],
\end{equation}
by referring the actual (true) local densities to the packing density $n_M$. Therefore, $f_1(t,\bx,\bv,u)d\bx d\bv du$ (respectively $f_2(t,\bx,\bv,u)d\bx d\bv du$) denotes the fraction of the \textit{walkers} (respectively of the \textit{leaders}), at time $t$, in the elementary volume $[\bx, \bx + d\bx] \times [\bv, \bv+d\bv] \times [u, u + du]$.

The macroscopic quantities are still defined by Eqs. \eqref{mac-1}-\eqref{mac-2}, in particular the \textit{initial numbers of walkers and leaders} are defined by
\begin{equation}
N_{i0} = \int_\Omega \int_0^1 \int_0^{2\pi}  \int_0^1 f_{i}(t=0,\,\bx,\,v, \theta, u)\,v dv\, d\theta\,du,\, d\bx \quad i=1,2.
\end{equation}
In addition, we introduce the following parameter
\begin{equation}
\sigma(\bx) = \frac{N_{20}(\bx)}{N_{10}(\bx)}\va
\end{equation}
which measures the presence of \textit{leaders} over the \textit{walkers}. In general, it is supposed that $\sigma$ is a small number  with respect to one.

The mathematical structure is obtained within the general framework given by Eq.(\ref{StructureII}), for a system whose state is described by the distribution functions $f_i = f_i(t, \bx, \bv, u)$, $i = 1, 2$,
\begin{equation}
\begin{cases}
\left(\p_t  +  \bv \cdot \p_\bx \right)\, f_1(t, \bx, \bv, u) = P_1[\f,f_1](t, \bx, \bv, u),\vspace{0.3cm}\\
\left(\p_t  +  \bv \cdot \p_\bx \right)\, f_2(t, \bx, \bv, u) = P_2[\f,f_2](t, \bx, \bv, u),
\end{cases}
\end{equation}
where
\begin{eqnarray}
&& P_1[\f,f_1]= \sum_{k=1}^2 \,  \eta_0 \,\int_{D^2}  \cA_{1k}[\f](\bv_* \to \bv, u_* \to u|\bv_*,  \bv^*, u_*, u^*;\Sigma) \nonumber \\
 && \hskip3truecm \times \, f_1(t, \bx, \bv_*, u_*) f_k(t, \bx,  \bv^*, u^*)\,d\bv_*\,d\bv^* \, du_*\, du^*, \nonumber \\
 && \hskip2truecm -  \eta_0 \, f_{1}(t, \bx, \bv,u)  \sum_{k=1}^2 \int_{D}  f_{k}(t, \bx,  \bv^*, u^*)\, d\bv^* \, du^*,
\end{eqnarray}
and
\begin{eqnarray}
&& P_2[\f,f_2]= \eta_0 \, \sum_{k=1}^2 \,  \int_{D^2}  \cA_{2k}[\f](\bv_* \to \bv, u_* \to u|\bv_*,  \bv^*, u_*, u^*;\Sigma) \nonumber \\
 && \hskip3truecm \times \, f_2(t, \bx, \bv_*, u_*) f_k(t, \bx,  \bv^*, u^*)\,d\bv_*\,d\bv^* \, du_*\, du^*, \nonumber \\
 && \hskip2truecm -  \eta_0\,f_{2}(t, \bx, \bv,u)  \sum_{k=1}^2 \int_{D}  f_{k}(t, \bx,  \bv^*, u^*)\, d\bv^* \, du^*,
\end{eqnarray}
where $\eta_0$ is a parameter which describes the frequency of interactions.

The derivation of the mathematical model is obtained by particularizing the interaction terms $\cA$. More precisely the transition probability density, as in Eq.~(14), is  factorized as follows:
\begin{equation}\label{factorization_A}
 \cA_{ik}(\bv_* \to \bv, u_* \to u) = \cA_{ik}^u(u^* \to u)\times \cA_{ik}^\theta (\theta_* \to \theta) \times \cA_{ik}^v(v_* \to v),
\end{equation}
where the terms $\cA_{ik}^u$, $\cA_{ik}^\theta$ and $\cA_{ik}^v$ correspond, respectively, to the dynamics of the emotional state, of the  selection of the walking direction and  of the walking speed. The table below, where only the dependence on $u$ has been indicated, summarizes  their expressions. 

\vskip.5cm
\begin{center}
\begin{tabular}{|c|l|}
  \hline
  \rule{0pt}{4.ex} {\bf Interaction} & {\bf \hspace{2.55cm} Probability transition }\\[2.ex]
  \hline
  &  \rule{0pt}{4.ex} $ \mathcal{A}_{11}(\bv_* \to \bv, u_* \to u| \bx, \bv_*, \bv^*, u_*, u^*)$
  \\
  (\textit{I-WW}) & \rule{0pt}{3.5ex} $ \hspace{0.25cm}= \big(\delta \big(u - \ve(u^* - u_*)(1 - u_*) \big) \times \delta \left(\theta[u, \cdot] -\theta_* \right) \times \delta \left(v[u, \cdot] -v_* \right)$\\[2.ex]
  \hline
   &  \rule{0pt}{4.ex} $\mathcal{A}_{12}(\bv_* \to \bv, u_* \to u| \bx, \bv_*, \bv^*, u_*, u^*)$\\
  (\textit{I-WL}) & $\hspace{0.25cm} = \delta( u - u_* + \varepsilon (u_* - u_0)) \times \delta \left(\theta[u, \cdot] -\theta_* \right) \times \delta \left(v[u, \cdot] -v_* \right) $\\[2.ex]
    \hline
  &  \rule{0pt}{4.ex} $\mathcal{A}_{21}(\bv_* \to \bv, u_* \to u| \bx, \bv_*, \bv^*, u_*, u^*)$\\
  (\textit{I-LW}) & $\hspace{0.25cm} = \delta( u - u_* + \varepsilon (u_* - u_0)) \times \delta \left(\theta[u, \cdot] -\theta_* \right) \times \delta \left(v[u, \cdot] -v_* \right) $\\[2.ex]
  \hline
  &  \rule{0pt}{4.ex}(\textit{I-LL}) $\mathcal{A}_{22}(\bv_* \to \bv, u_0 \to u| \bx, \bv_*, \bv^*, u_0, u^*)$ \\
  (\textit{I-LL}) & $\hspace{0.25cm} = \delta(u_* - u_0) \times \delta \left(\theta[u_0, \cdot] -\theta_* \right) \times \delta \left(v[u_0, \cdot] -v_* \right)$\\[2.ex]
  \hline
\end{tabular}
\end{center}
\vskip.5cm

This modeling result has been obtained under the following assumptions:
\begin{enumerate}

\item The activity, at $t=0$, is homogeneously distributed  with value $u_0$ both for leaders and walkers;

 \vskip.1cm \item \textit{Walker-walker interactions:} The activity is not modified by the presence of leaders, so that the probability density $\cA_{11}^u$ is still given by Eqs.~\eqref{Au-1},\eqref{Au-2};

  \vskip.1cm \item \textit{Walker-leader interactions:} The activity of walkers has a trend toward the activity of the leaders:
\begin{equation}
\mathcal{A}^u_{12}(u_* \to u|u_*, u^*) = \delta(u - u_* + \varepsilon (u_* - u_0)),
\end{equation}
subsequently the dynamics of $\theta$ and $v$ follows the same rules as in the preceding item;

 \vskip.1cm \item \textit{Leader-walker-leader and leader-leader interactions:}  The activity is not modified by both these interactions:
\begin{equation}
\mathcal{A}^u_{21}(u_0 \to u|u_0) = \delta(u - u_0), \quad \hbox{and} \quad  \mathcal{A}^u_{22}(u_0 \to u|u_0) = \delta( u - u_0),
\end{equation}
therefore the dynamics of $\theta$ and $v$ follows the same rules of the walker, but with $u=u_0$.

\end{enumerate}

\section{A Case Study and Simulations}

This section presents some simulations developed to test the predictive ability of the models proposed in Section 4. The mathematical problem that generates these simulations needs the statement of initial conditions $f(t=0, \bx, \bv, u)$ and boundary conditions which are necessary although the walking strategy attempts to avoid the encounter with the walls. In fact, some of the walkers, however viewed as active particles, might reach, in probability, the wall, then an appropriate reflection model at the boundary must be given.

The Boltzmann-like structure of the equation requires boundary conditions analogous to those used by the fundamental model of the classical kinetic theory. In more detail, we suppose that interaction with the wall modifies only
the direction of velocity, after the dynamics follows the same rules already stated in Section 3. Accordingly, the statement of boundary conditions can be given as follows:
\begin{equation}
\label{bconditions}
f^r(t, \bx, \theta_r,u) = \frac{|\bv_i\cdot \bn|}{|\bv_r\cdot \bn|} \int \, R( \theta_i \to  \theta_r)\, f^i(t, \bx, \theta_i, u) \,d\theta_i',
\end{equation}
where $f^r$ and $f^i$ denote, respectively, the distribution function after and before interactions with the wall, while $\theta_i$ and  $\theta_r$ denote the velocity directions before and after the interaction. These directions are, respectively  such that $\bv \cdot \bn \leq 0$ and $\bv \cdot \bn \geq 0$, where $\bn$ is the unit vector orthogonal do the wall and directed inside the domain.

Bearing all above in mind, we can now define the specific problem we will address the simulation to. The main features of the case study are the following:

\begin{itemize}

\item The crowd  is constituted by  Two groups of people move in opposite directions in a  rectangular venue of $20\,m \, \times 5\,m$;

\vskip.1cm \item The group on the left is composed of $40$ people uniformly distributed in a rectangular area $4\,m\times4\,m$  with the initial emotional state set to $u\simeq0.4$ while the group on the right is composed of $20$ people uniformly distributed in a rectangular area of $4\,m\times2\,m$ with an higher level of stressful condition, namely $u\simeq0.8$;

\vskip.1cm \item The speed $\xi$ is also homogeneously  distributed over all walkers at a value $\xi_0 \cong u_0$;

\vskip.1cm \item When the two groups physically interact, a mixing of stress conditions appears, which modifies the walking dynamics which would occur in absence of social interaction.

\end{itemize}

The objective of simulations consists in understanding how social interactions modify the patterns of the flow and how high density patterns localize.
A quantity which worth to be computed is the \textit{mean density of the emotional state} 
\begin{equation}
  \bar{u}(t,\bx) = \frac{1}{ \rho(t,\bx)}\int f(t,\bx,\bv, u) u \,d\bv \, du.
 \end{equation}

\begin{figure}[t!]

      \begin{subfigure}{0.45\textwidth}
          \centering \includegraphics[height=2.0cm]{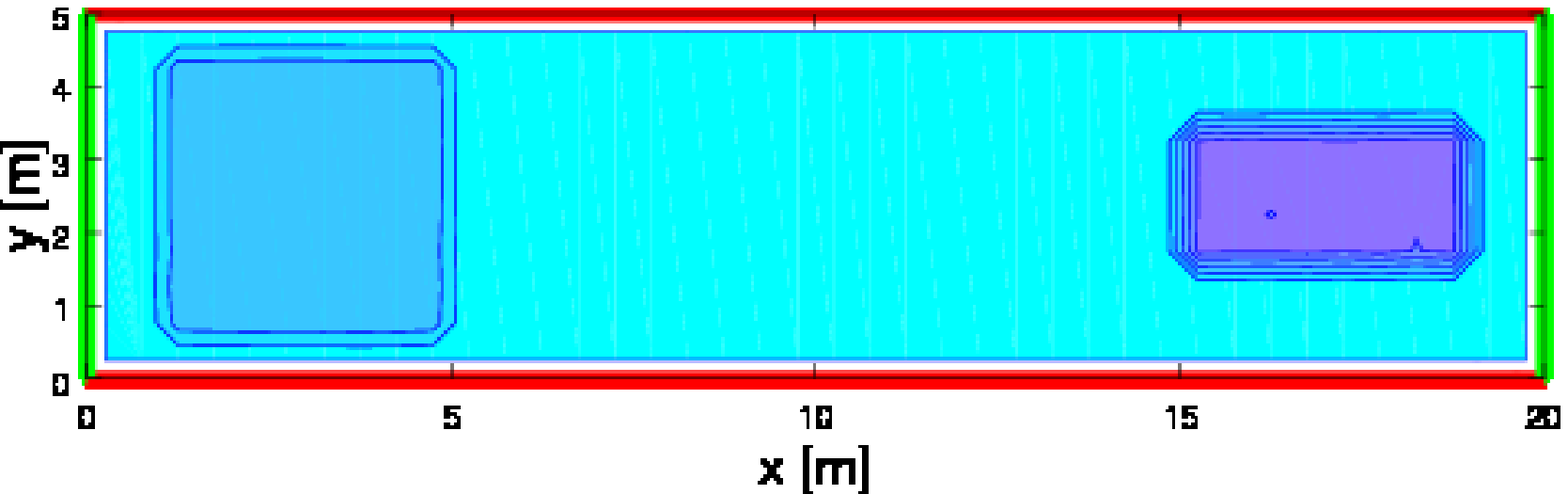}
          \subcaption{$t=0\,s$.}
    \end{subfigure}   \hfill
    \begin{subfigure}{0.45\textwidth}
          \centering \includegraphics[height=2.0cm]{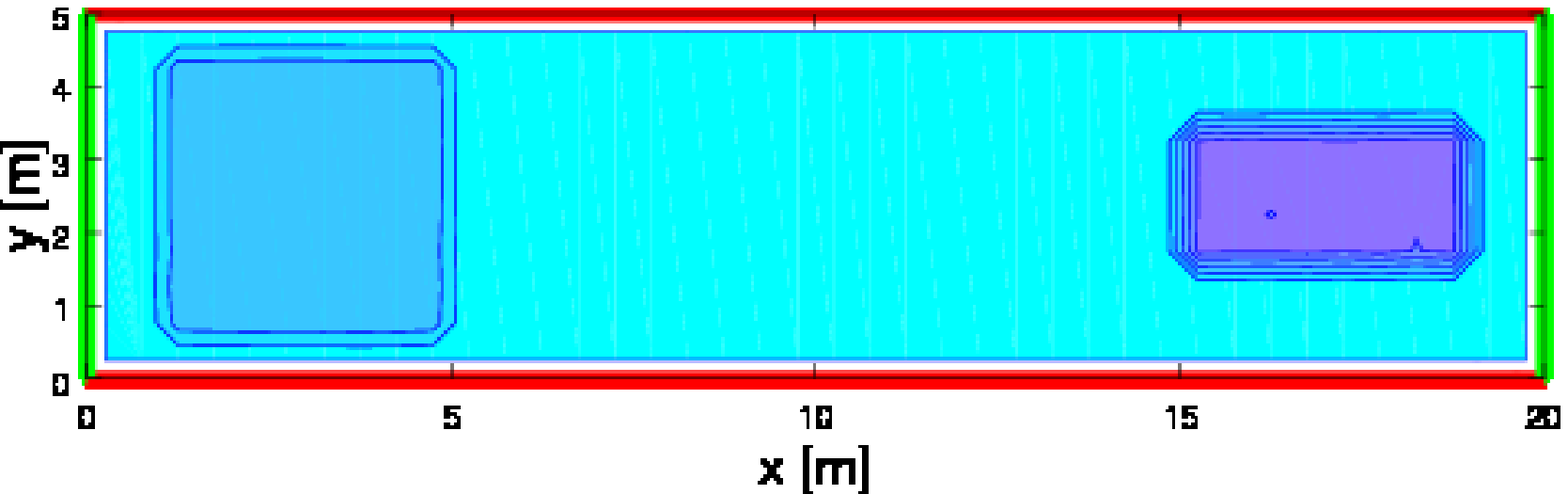}
          \subcaption{$t=0\,s$.}
    \end{subfigure}

     \vspace{0.5cm}

     \begin{subfigure}[b]{0.45\textwidth}
      \begin{center}
         \includegraphics[height=2.0cm]{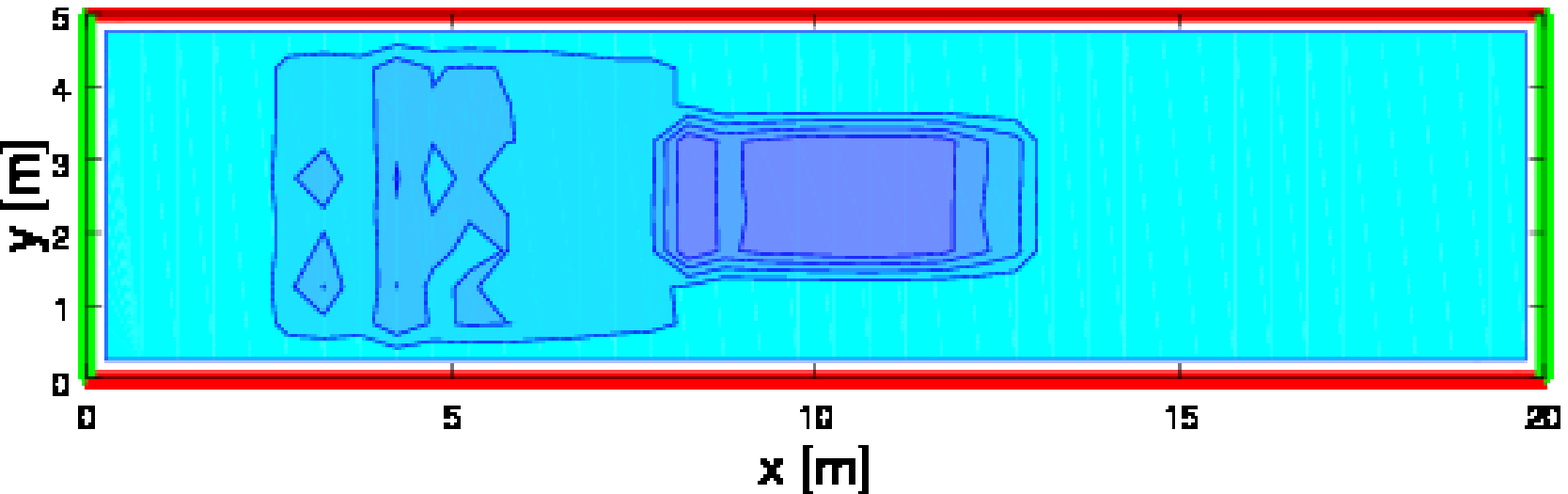}
         \subcaption{$t=6\,s$.}
      \end{center}
   \end{subfigure}   \hfill
   \begin{subfigure}[b]{0.45\textwidth}
      \begin{center}
         \includegraphics[height=2.0cm]{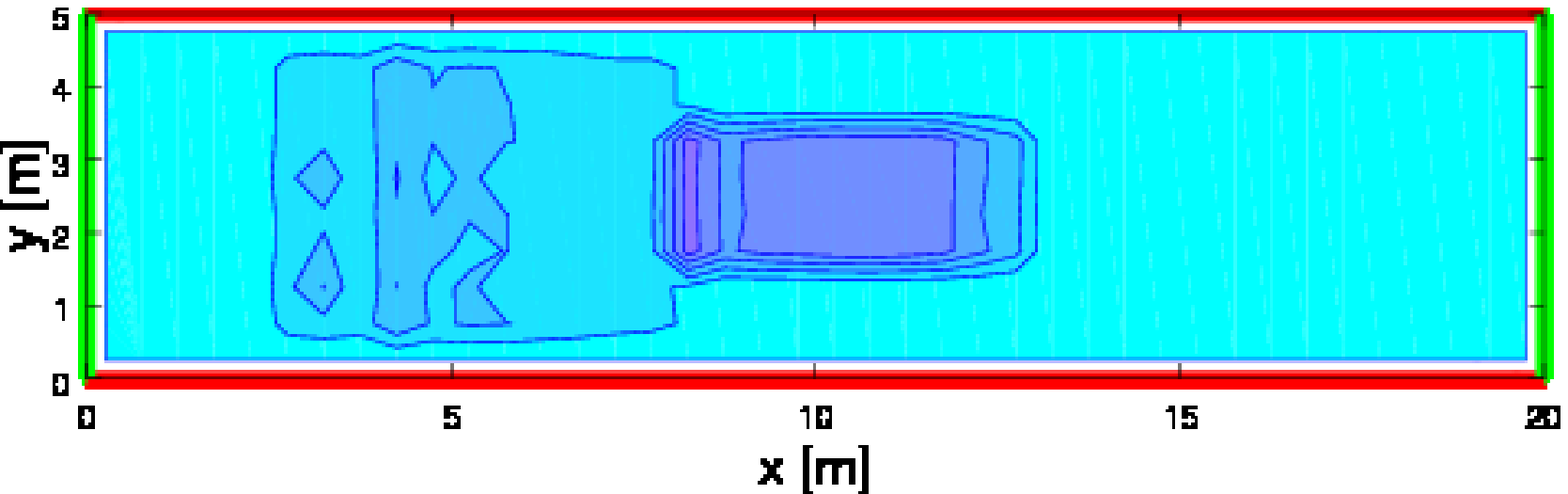}
         \subcaption{$t=6\,s$.}
      \end{center}
   \end{subfigure}

    \vspace{0.5cm}

     \begin{subfigure}[b]{0.45\textwidth}
      \begin{center}
         \includegraphics[height=2.0cm]{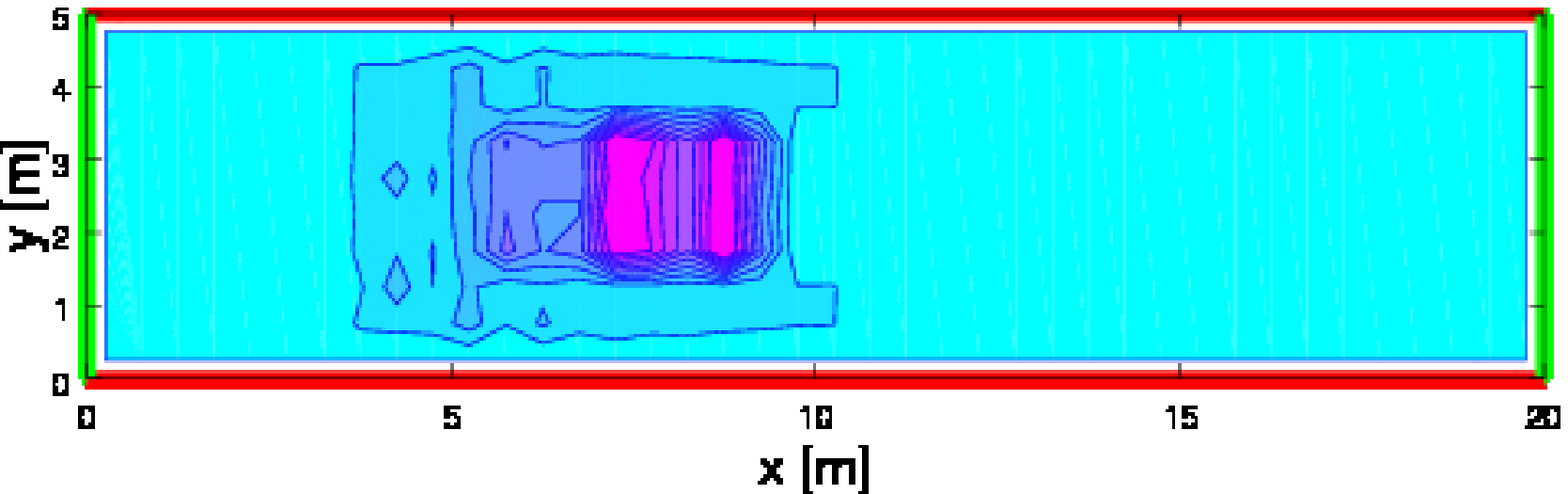}
         \subcaption{$t=10\,s$.}
      \end{center}
   \end{subfigure}   \hfill
   \begin{subfigure}[b]{0.45\textwidth}
      \begin{center}
         \includegraphics[height=2.0cm]{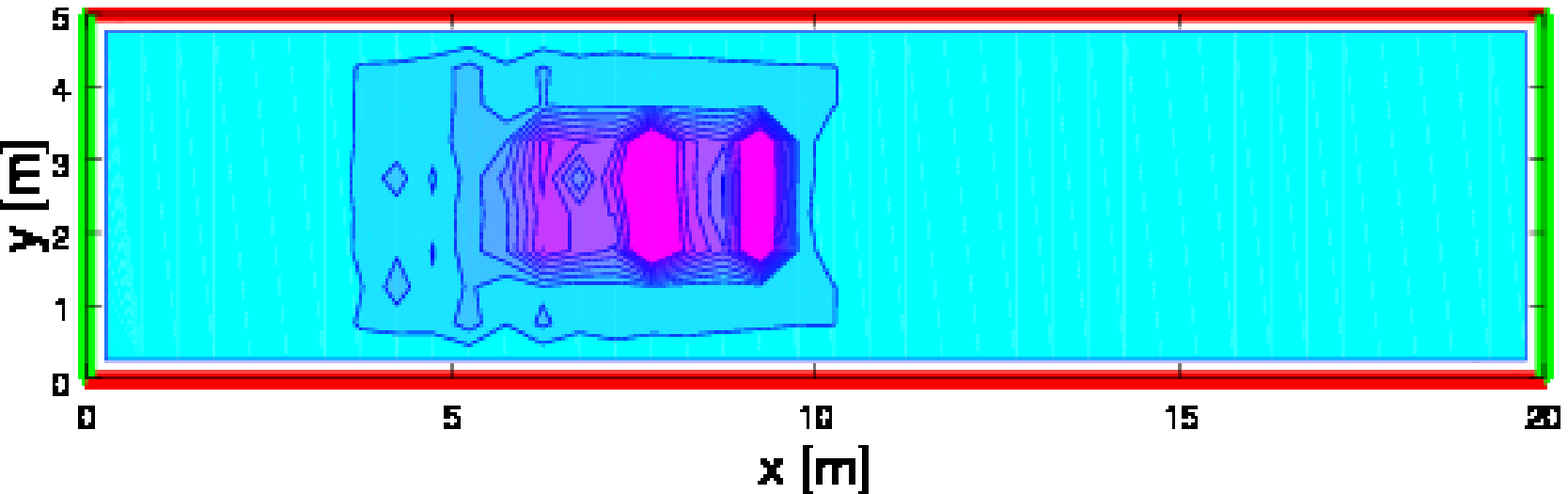}
         \subcaption{$t=10\,s$.}
      \end{center}
   \end{subfigure}

 \vspace{0.5cm}

  \caption{Density contour plot of the mean density of the emotional state, $\rho \bar{u}$,
           without (left panels) and with (right panels) social interactions.}
 \label{fig:withwithout_1}
 \end{figure}

\begin{figure}[t!]

      \begin{subfigure}[b]{0.45\textwidth}
      \begin{center}
         \includegraphics[height=2.0cm]{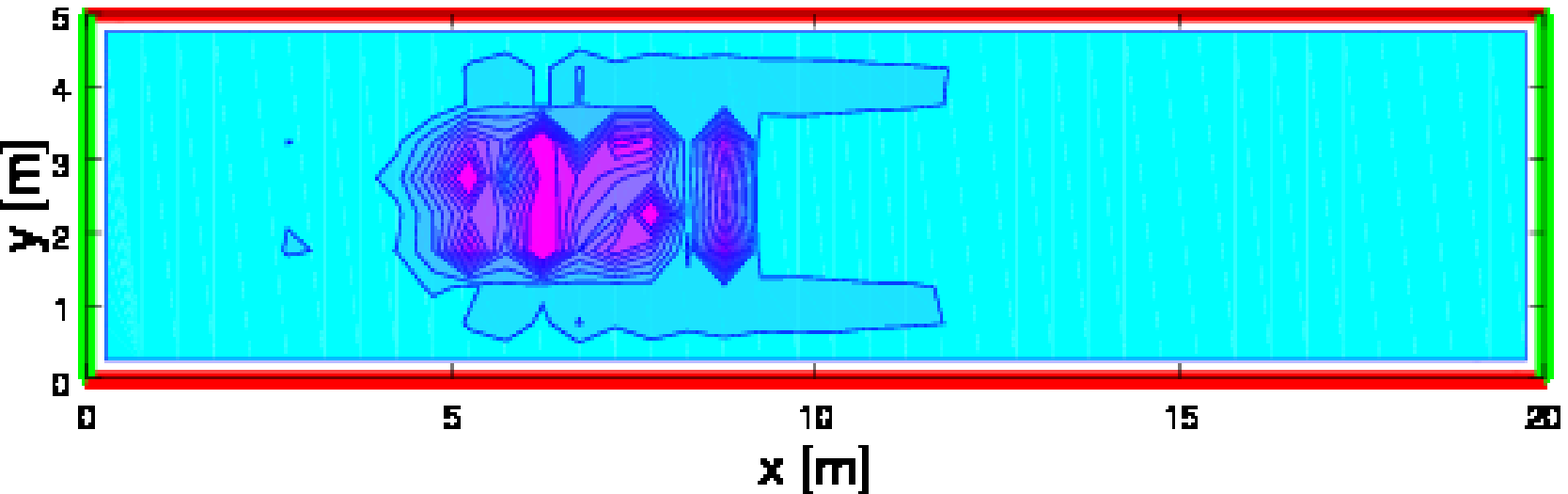}
         \subcaption{$t=14\,s$.}
      \end{center}
   \end{subfigure}   \hfill
   \begin{subfigure}[b]{0.45\textwidth}
      \begin{center}
         \includegraphics[height=2.0cm]{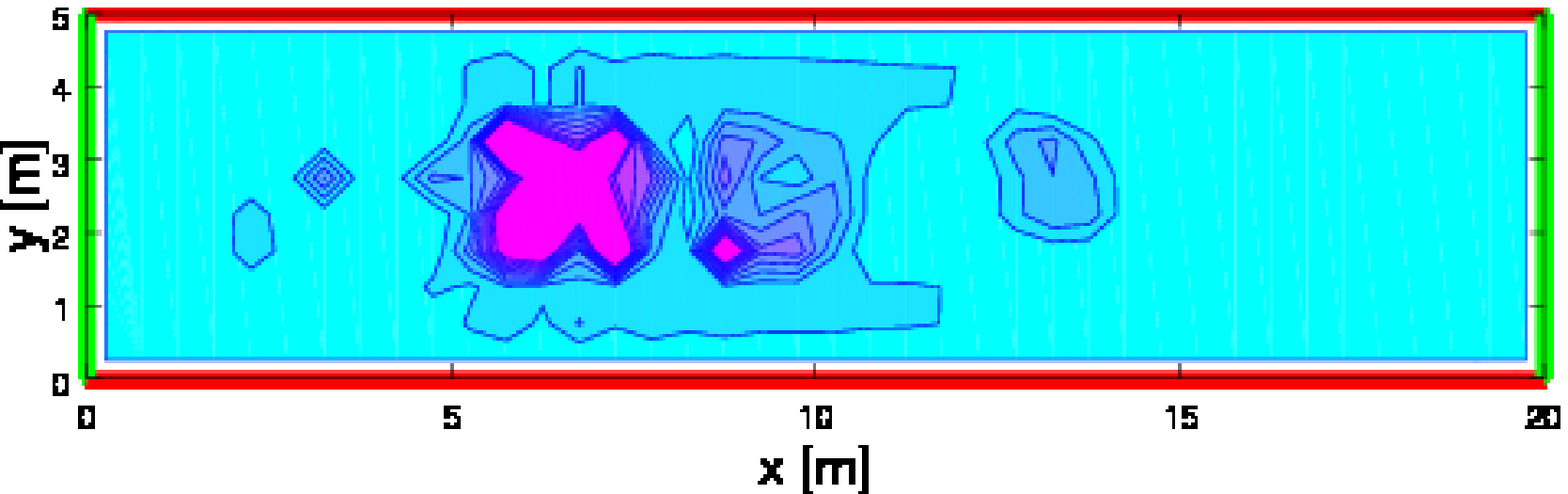}
         \subcaption{$t=14\,s$.}
      \end{center}
   \end{subfigure} 
   
   \vspace{0.5cm}

     \begin{subfigure}[b]{0.45\textwidth}
      \begin{center}
         \includegraphics[height=2.0cm]{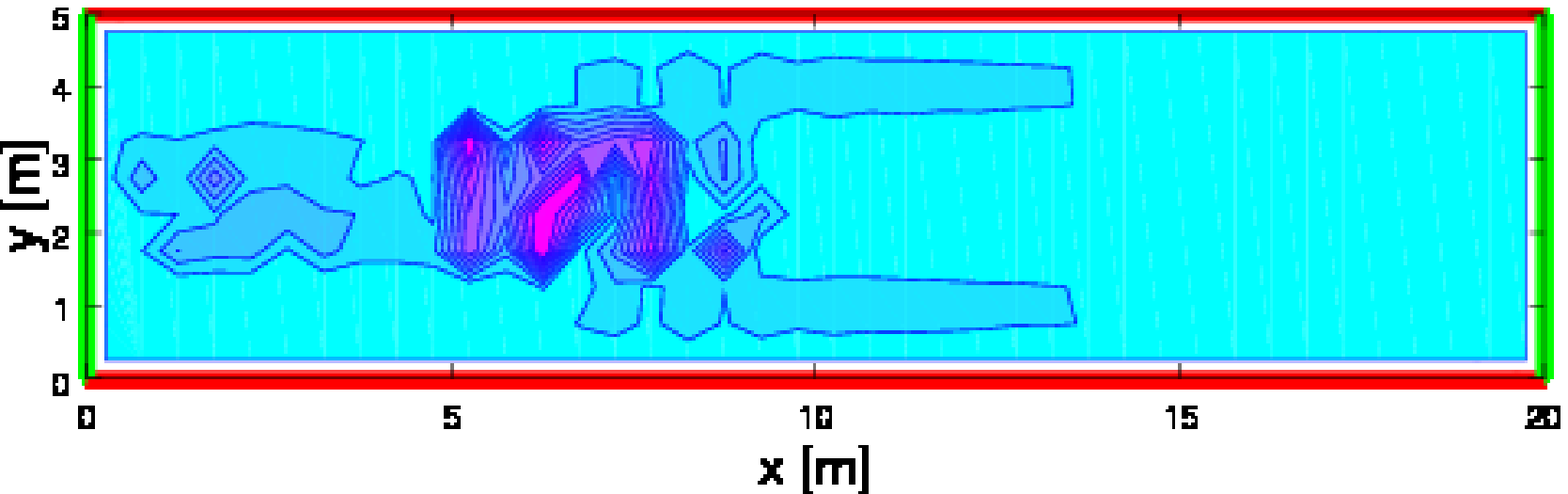}
         \subcaption{$t=18\,s$.}
      \end{center}
   \end{subfigure}   \hfill
   \begin{subfigure}[b]{0.45\textwidth}
      \begin{center}
         \includegraphics[height=2.0cm]{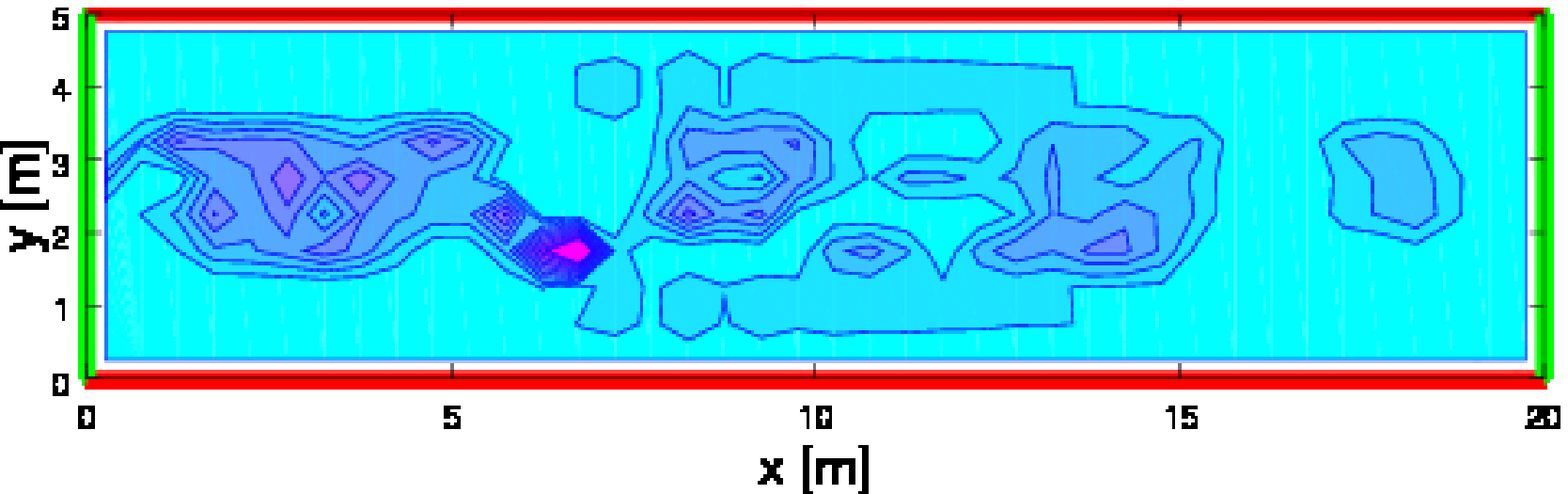}
         \subcaption{$t=18\,s$.}
      \end{center}
   \end{subfigure}

     \vspace{0.5cm}

     \begin{subfigure}[b]{0.45\textwidth}
      \begin{center}
         \includegraphics[height=2.0cm]{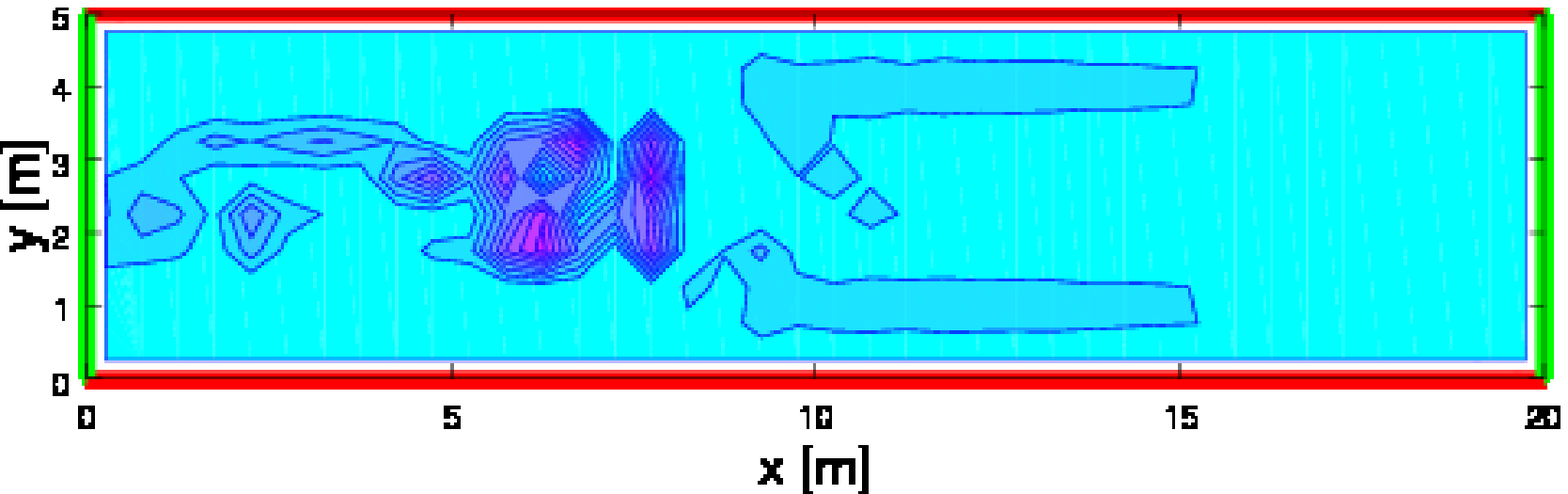}
         \subcaption{$t=22\,s$.}
      \end{center}
   \end{subfigure}   \hfill
   \begin{subfigure}[b]{0.45\textwidth}
      \begin{center}
         \includegraphics[height=2.0cm]{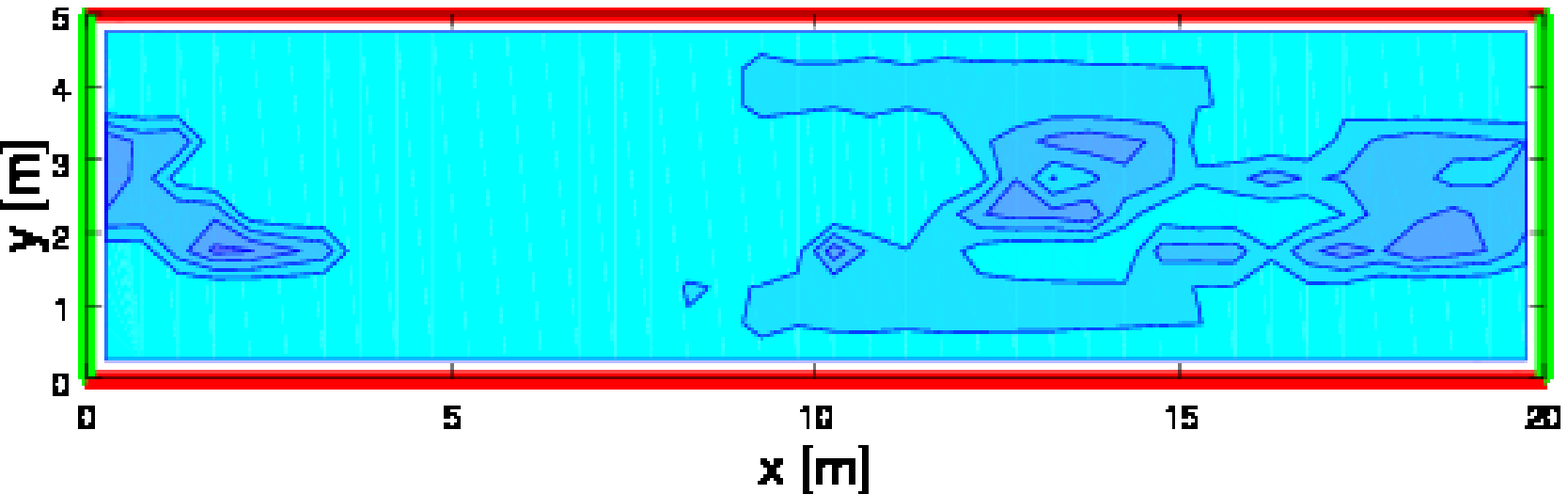}
         \subcaption{$t=22\,s$.}
      \end{center}
   \end{subfigure}

%

   \vspace{0.5cm}

  \caption{Density contour plot of the mean density of the emotional state, $\rho \bar{u}$,
           without (left panels) and with (right panels) social interactions}
 \label{fig:withwithout_2}
 \end{figure}

Simulations related to the case study under consideration are reported in Figures~\ref{fig:withwithout_1} and~\ref{fig:withwithout_2} which  show, respectively, the contour plots of the mean density of the emotional state with (right panels) and without (left panels) social
 interactions for different times. These figures put in evidence how the exchange of emotional states modifies the aforementioned patterns and, specifically, induces zones with high density concentration which, as it is known, can generate loss of safety conditions.

\newpage

\section{Critical Analysis and Research Perspectives}

A kinetic theory approach to the modeling of crowd dynamics in the presence of social phenomena, which can modify the rules of mechanical interactions, has been proposed in our paper. Two types of social dynamics have been specifically studied, namely the propagation of stress conditions and the role of leaders. The case study proposed in Section 5 has shown that stress conditions can induce important modifications in the overall dynamics and on the density patterns thus enhancing formation of overcrowded zones. The specific social dynamics phenomena studied in our paper have been motivated by situations, such as fire incidents or rapid evacuations, where safety problems can arise~\cite{[EPS02],[LL15],[RNC16],[WCM16]}.

The achievements presented in the preceding sections motivate a systematic computational analysis focused on a broader variety of case studies focusing specifically to enlarge the variety of social phenomena inserted in the model. As an example, one might consider even extreme situations, where antagonist groups contrast each other in a crowd. This type of developments can be definitely inserted into a possible research program which is strongly motivated by the security problems of our society.

Furthermore, we wish returning to the scaling problem, rapidly introduced in Sections 1 and 2, to propose a critical, as well as self-critical, analysis induced also by the achievements of our paper on the modeling human behaviors in crowds. In more detail, we observe that it would be useful introducing aspects of social behaviors also in the modeling at the microscopic and macroscopic scale. Afterwards, a critical analysis can be developed  to enlighten advantages and withdraws of the selection of a certain scale with respect to the others.

This type of analysis should not hide the conceptual link which joins the modeling approach at the different scales. In fact a detailed analysis of individual based interactions (microscopic scale)  should implement the derivation of kinetic type models (mesoscopic scale), while hydrodynamic models (macroscopic scale) should be derived from the underlying description delivered kinetic type models by asymptotic methods where a small parameter corresponding to the distances between individuals is let to tend to zero.  Often models are derived independently at each scale, which prevents a real multiscale approach.

Some achievements have already been obtained on the derivation of macroscopic equations from the kinetic type description for crowds in unbounded domains~\cite{[BB15]} by an approach which has some analogy with that developed for vehicular traffic~\cite{[BBNS14]}. However, applied mathematicians might still investigate how the structure of macroscopic models is modified by social behaviors. This challenging topic might be addressed even to the relatively simpler problem of vehicular traffic where individual behaviors are taken into account~\cite{[BDF17]}.

 Finally, let us state that the ``important'' objective, according to our own bias, is the development of a systems approach to crowd dynamics, where models derived at the three different scales might coexist in complex venues where the local number density from rarefied to high number density. This objective induces the derivation of models at the microscopic scale consistent with models at the macroscopic scale with the intermediate description offered by the kinetic theory approach.

\textit{E-mail address:} \href{mailto:nicola.bellomo@polito.it}{nicola.bellomo@polito.it} 

\textit{E-mail address:} \href{mailto:L.Gibelli@warwick.ac.uk}{L.Gibelli@warwick.ac.uk} 

\textit{E-mail address:} \href{mailto:outada@ljll.math.upmc.fr}{outada@ljll.math.upmc.fr} 

\begin{thebibliography}{99}

\bibitem{[ABG16]}
\newblock G.~Ajmone Marsan, N.~Bellomo, and L.~Gibelli,
\newblock Stochastic evolutionary differential games toward a systems theory of behavioral social dynamics,
\newblock {\em Math. Models Methods Appl. Sci.}, {\textbf{26(6)} (2016), 1051--1093.

\bibitem{[ARI01]}
    \newblock V.V.~Aristov,
    \newblock ``Direct Methods for Solving the Boltzmann Equation and Study of Nonequilibrium Flows'',
    \newblock Springer-Verlag, New York, 2001.

\bibitem{[BFG15]}
    \newblock P.~Barbante, A.~Frezzotti, and L.~Gibelli,
    \newblock A kinetic theory description of liquid menisci at the microscale,
    \newblock {\em Kinet. Relat. Models}, \textbf{8(2)} (2015), 235--254.


\bibitem{[BB15]}
\newblock N.~Bellomo and A.~Bellouquid,
\newblock On multiscale models of pedestrian crowds from mesoscopic to macroscopic,
\newblock {\em Comm. Math. Sciences}, \textbf{13(7)} (2015), 1649--1664.

\bibitem{[BBNS14]}
\newblock N.~Bellomo, A.~Bellouquid, J.~Nieto, and J.~Soler,
\newblock On the multiscale modeling of vehicular traffic: From kinetic to hydrodynamics,
\newblock {\em Discr. Cont. Dyn. Syst. Series B}, \textbf{19}, 1869--1888, (2014).


 \bibitem{[BBK13]}
 \newblock N.~Bellomo, A.~Bellouquid, and D.~Knopoff,
     \newblock     From the micro-scale to collective crowd dynamics,
        \newblock  {\em Multiscale Model. Sim.},  \textbf{11} (2013), 943--963.


\bibitem{[BCG16]}
\newblock N.~Bellomo, D.~Clark, L.~Gibelli, P.~Townsend, and  B.J.~Vreugdenhil,
\newblock Human behaviours in evacuation crowd dynamics: From modelling to big data toward crisis management.
\newblock {\em Phys. Life Rev.}, \textbf{18} (2016), 1--21.


\bibitem{[BKS13]}
\newblock N.~Bellomo, D.~Knopoff, and J.~Soler,
\newblock On the difficult interplay between life, ``complexity'', and mathematical sciences.
\newblock {\em Math. Models Methods Appl. Sci.}, \textbf{23} (2013), 1861--1913.


\bibitem{[BG15]}
\newblock N.~Bellomo and L.~Gibelli,
\newblock Toward a behavioral-social dynamics of pedestrian crowds,
\newblock \emph{Math. Models Methods Appl. Sci.}, \textbf{25} (2015), 2417--2437.

\bibitem{[BG16]}
\newblock N.~Bellomo and L.~Gibelli,
\newblock Behavioral crowds: Modeling and Monte Carlo simulations toward validation.
    \newblock Comp.\& Fluids, \textbf{141} (2016), 13--21.


\bibitem{[BIR94]} G.A.~Bird,
    \newblock  ``Molecular Gas Dynamics and the Direct Simulation of Gas Flows'',
    \newblock Oxford University Press, (1994).
    
\bibitem{[BDF17]}
    \newblock D.~Burini, S.~De Lillo, and G.~Fioriti,
    \newblock Influence of drivers ability in a discrete vehicular traffic model
    \newblock  {\em Int. J. Modern Phys.},  \textbf{28(3)}, (2017), 1750030.
    

\bibitem{[BDG16]}
\newblock D.~Burini, S.~De Lillo~S., and L.~Gibelli,
\newblock Stochastic differential ``nonlinear''  games modeling collective learning dynamics,
\newblock {\em Phys. Life Rev.}, \textbf{16(1)} (2016), 123--139.

\bibitem{[BDG16B]}
\newblock  D.~Burini, S.~De Lillo~S., and L.~Gibelli,
\newblock   Learning dynamics towards modeling living systems. Reply to comments on  ``Stochastic differential ``nonlinear''  games modeling collective learning dynamics'',
\newblock {\em Phys. Life Rev.}, \textbf{16(1)} (2016), 158--162.


\bibitem{[CIP93]}
\newblock C.~Cercignani, R.~Illner, and M.~Pulvirenti,
\newblock  ``The Mathematical Theory of  Diluted Gas'',
\newblock Springer, Heidelberg, New York, (1993.

\bibitem{[CMV15]}
\newblock A.~Corbetta, A.~Mountean, and K.~Vafayi,
\newblock Parameter estimation of social forces in pedestrian dynamics models via probabilistic method,
\newblock {\em Math. Biosci. Eng.}, {\bf 12} (2015), 337--356.


\bibitem{[CPT14]}
\newblock E.~Cristiani, B.~Piccoli, and A.~Tosin,
\newblock  ``Multiscale Modeling of Pedestrian Dynamics'',
  \newblock Springer, (2014).

  \bibitem{[DAM13]}
\newblock  P.~Degond, C.~Appert-Rolland, M.~Moussa\"id, J.~Pettr\'e, and G.~Theraulaz,
\newblock A Hierarchy of Heuristic-Based Models of Crowd Dynamics,
\newblock {\em J. Stat. Phys.}, {\bf 152} (2013), 1033--1068.

\bibitem{[DLM17]}
\newblock P.~Degond,  J.-G.~Liu, S.~Merino-Aceituno, and T.~Tardiveau,
\newblock Continuum dynamics of the intention field under weakly cohesive social interaction,
\newblock {\em Math. Models Methods Appl. Sci.}, \textbf{27} (2017), 159--182.


\bibitem{[EPS02]}
\newblock J.M.~Epstein,
\newblock Modeling civil violence: An agent based computational approach,
\newblock {\em Proc.  Nat. Acad. Sci.}, \textbf{99} (2002),  7243--7250.


\bibitem{[FPA11]}
\newblock R.F.~Fahy, G.~Proulx, and L.~Aiman,
\newblock Panic or not in fire: Clarifyng the misconception,
\newblock {\em Fire Material}, Wiley on Line Librery, DOI: 10.1002/fam.1083.



\bibitem{[HEL01]}
\newblock D.~Helbing,
\newblock Traffic and related self-driven many-particle systems,
\newblock {\em Rev. Modern Phys.}, \textbf{73} (2001),  1067--1141.



\bibitem{[HFV00]}
    \newblock D.~Helbing, I.~Farkas, and  T.~Vicsek,
    \newblock Simulating dynamical feature of escape panic.
    \newblock {\em Nature}, \textbf{407} (2000),  487--490.


\bibitem{[HJ09]}
    \newblock D.~Helbing and A.~Johansson,
    \newblock Pedestrian crowd and evacuation dynamics,
    \newblock {\em Enciclopedia of Complexity and System Science}, (2009), 6476--6495.


\bibitem{[HJA07]}
    \newblock D.~Helbing, A.~Johansson,  and H.Z.~Al-Abideen,
    \newblock Dynamics of crowd disasters: An empirical study,
    \newblock {\em Phys. Rev. E},  \textbf{75} (2007), paper no.~046109.

\bibitem{[HUG03]}
\newblock R.L.~Hughes,
\newblock The flow of human crowds,
\newblock  {\em Annu. Rev. Fluid Mech.}, \textbf{35} (2003), 169--182.

\bibitem{[KIN13]}
\newblock M.~Kinateder \textit{et al.},
\newblock Human behaviour in severe tunnel accidents: Effects of information and behavioural training,
\newblock {\em Transp. Res. Part F: Traffic Psychology and Behaviour}, \textbf{17} (2013), 20--32.


\bibitem{[LEB02]}
\newblock G.~Le Bon,
\newblock ``The Crowd. A Study of the Popular Mind'',
\newblock Dover Pub., (2002).


\bibitem{[LL15]}
\newblock J.~Lin and T.A.~Luckas,
\newblock A particle swarm optimization model of emergency airplane evacuation with emotion,
\newblock  {\em Net. Het. Media}, \textbf{10} (2015), 631--646.

\bibitem{[MMGM17]}
\newblock S.~Motsch, M.~Moussa\"id, E.-G.~Guillot, M.~Moreau, J.~Pettr\'e, G.~Theraulaz,  C.~Appert-Rolland, and P.~Degond,
\newblock Forecasting crowd dynamics through coarse-grained data analysis,
\newblock bioRxiv preprint first posted online Aug. 13, 2017; doi: http://dx.doi.org/10.1101/175760.


\bibitem{[MHG09]}
    \newblock Moussa\"id~M., Helbing~D., Garnier~S., Johansson~A., Combe~M., and Theraulaz~G.,
    \newblock Experimental study of the behavioural mechanisms underlying  self-organization in human crowds,
    \newblock {\em Proc. Roy. Soc. B},  \textbf{276} (2009), 2755--2762.
    

\bibitem{[MT11]}
\newblock Moussa\"id~M. and Theraulaz~G.,
\newblock Comment les pi\'etons marchent dans la foule.
\newblock {\em La Recherche},   \textbf{450} (2011), 56--59.


\bibitem{[PT14]}
    \newblock L.~Pareschi and G.~Toscani,
    \newblock  ``Interacting Multiagent Systems: Kinetic Equations and Monte Carlo Methods'',
    \newblock Oxford University Press, Oxford, (2014).


\bibitem{[RON15]}
\newblock E.~Ronchi,
\newblock Disaster management: Design buildings for rapid evacuation
\newblock {\em Nature},  \textbf{528)} (2015), 333.

\bibitem{[RGP13]}
\newblock E.~Ronchi,  S.M.V.~Gwynne, D.A.~Purser, and P.~Colonna,
\newblock Representation of the Impact of Smoke on Agent Walking Speeds in Evacuation Models
\newblock  {\em Safety Science}, \textbf{52} (2013), 28--36.


\bibitem{[RKN16]}
\newblock E.~Ronchi,  E.D.~Kuligowski, D.~Nilsson, R.D.~Peacock, and P.A.~Reneke,
\newblock Assessing the verification and validation of building fire evacuation models
\newblock {\em Fire Technology}, \textbf{52(1)} (2016), 197--219.



\bibitem{[RNC16]}
    \newblock F.~Ronchi, F.~Nieto Uriz, X.~Criel, and P.~Reilly,
    \newblock Modelling large-scale evacuation of music festival.
    \newblock  {\em Fire Safety}, \textbf{5} (2016), 11--19.


\bibitem{[RRP16]}
\newblock E.~Ronchi, P.A.~Reneke, and R.D.~Peacock,
\newblock A conceptual fatigue-motivation model to represent pedestrian movement during stair evacuation,
\newblock {\em Appl. Math. Mod.},  \textbf{40(7-8)} (2016), 4380--4396.


\bibitem{[SW05]}
\newblock D.~Schweingruber and R.T.~Wohlstein,
The madding crowd goes to school: myths about crowds in introductory sociology textbooks.
 {\em Teaching Sociology Compass},  \textbf{33} (2005), 136-15}.
 
\bibitem{[TAL07]}
\newblock N.N.~Taleb,
\newblock ``The Black Swan: The Impact of the Highly Improbable'',
\newblock New York City: Random House, (2007).

 \bibitem{[WCM16]}
\newblock N.~Wijermans, C.~Conrado, M.~van Steen, C.~Martella, and J.L.~Li,
\newblock A landscape of crowd management support: An integrative approach,
\newblock  {\em Safety Science}, \textbf{86} (2016),  142--164.

\bibitem{[WSB17]}
    \newblock L.~Wang, M.~Short, and A.L.~Bertozzi,
    \newblock Efficient numerical methods for multiscale crowd dynamics with emotional contagion,
    \newblock  \emph{Math. Models Methods Appl. Sci.}, \textbf{27(1)} (2017),  205--230.


\bibitem{[WJJ13]}
\newblock N.~Wijermans, R.~Jorna, W.~Jager, T.~van Vliet, and O.M.J.~Adang,
\newblock CROSS: Modelling crowd behaviour with social-cognitive agents,
\newblock  {\em Journal of Artificial Societies and Social Simulation}, \textbf{16(4)} (2013), 1--14.

\bibitem{[WIN12]}
\newblock H.~Winter,
\newblock ``Modelling Crowd Dynamics During Evacuation Situations Using Simulation'',
\newblock Lancaster University, STOR-601, Project 2012.
 \end{thebibliography}
\end{document}